\documentclass[twocolumn,08 pt,amsmath,amssymb,aps,fleqn]{revtex4-1}
\usepackage[hidelinks]{hyperref}
\usepackage[english]{babel}
\usepackage{mathtools}
\usepackage{tabularx}
\usepackage{multirow}
\usepackage{comment}
\usepackage{graphicx}
\usepackage{color}
\usepackage{float}
\usepackage{soul,cancel}
\begin{document}

\preprint{APS/123-QED}
\title{Single-Particle Spectroscopy and Cross-Barrier Dynamics in Inhomogeneous Luttinger Liquids via Unified Field Bosonization Technique}
\author{Soundarya P$^{ 1}$, Venkata Suryanarayana M$^{ 1}$ and Joy Prakash Das$^{ 2*}$}
\affiliation{
	$^{\it 1}$Department of Physics, National Institute of Technology Tiruchirappalli, Tamil Nadu - 620015, India\\
	$^{\it 2}$Department of Physics, Assam Engineering College, Guwahati, Assam - 781013, India
		   }
\email{jpdas100@gmail.com}
\begin{abstract}
The Unified Field Bosonization Technique (UFBT) is employed to investigate the single-particle properties of inhomogeneous Luttinger liquids containing localized clusters of impurities in the presence of forward-scattering interactions. Within this framework, fermionic two-point correlation functions are used to derive the leading-order singular contributions to the dynamical density of states (DDOS) and the cross-barrier tunneling conductance. The results reveal distinct power-law behaviours of the DDOS near the impurity cluster and in the far-field regime. The tunneling conductance exhibits a power-law temperature dependence, decreasing with decreasing temperature for repulsive interactions and increasing for attractive interactions, while remaining temperature-independent in the noninteracting limit. The impurity cluster modifies the amplitudes and spatial structure of these observables, whereas their characteristic power-law exponents are determined solely by the interaction parameters. The formulation accommodates arbitrary impurity strengths without requiring perturbative expansions in the impurity or interaction strengths. The characteristic scaling behaviour emerges directly from the fermionic correlation functions, without an explicit renormalization-group (RG) treatment of the impurity strength, while the weak-interaction limit recovers the leading perturbative correction to the tunneling response. These findings establish UFBT as a unified framework for describing local spectral properties and cross-barrier transport in inhomogeneous Luttinger liquids.
\end{abstract}

\maketitle
\section{Introduction}

One-dimensional (1D) quantum systems occupy a distinctive position in many-body physics because interactions have a profound effect on their low-energy properties \cite{giamarchi2004quantum}. In contrast to higher-dimensional systems, where the Fermi-liquid picture provides a useful description of interacting fermions, the low-energy excitations of a generic interacting 1D system are collective in nature and are described by the Luttinger-liquid paradigm \cite{haldane1981luttinger}. The increasing experimental realization of effectively one-dimensional electronic systems, including carbon nanotubes \cite{bockrath1999luttinger,bockrath2001resonant,yao2000high} and semiconductor quantum wires \cite{auslaender2000experimental,yacoby1997magneto}, has further motivated the study of their correlation functions and experimentally accessible response properties.

Correlation functions provide the basic theoretical connection between the microscopic description of a many-body system and measurable quantities \cite{ge2024analytic}. In 1D interacting systems, bosonization provides a particularly powerful framework in which fermionic operators are represented in terms of bosonic collective fields \cite{von1998bosonization,giamarchi2004quantum}. For a homogeneous Luttinger liquid, this approach yields analytical expressions for a broad class of $N$-point correlation functions and has consequently been used to study spectral density and transport properties. The situation becomes substantially more intricate in the presence of impurities. Even a localized scatterer can have a non-perturbative influence on a 1D interacting system, since the impurity couples directly to the collective low-energy modes. The resulting competition between interactions and impurity scattering gives rise to characteristic phenomena such as impurity-induced density oscillations, suppression of the local density of states, resonant transmission, and strong renormalization of conductance \cite{das2019transport, das2020friedel}.

The special role of impurities in a Luttinger liquid was established particularly clearly by Kane and Fisher \cite{kane1992transport,kane1992resonant}. Their analysis showed that repulsive interactions enhance impurity strength at low energies, whereas attractive interactions render even strong impurities effectively weak. This physics, often described in terms of the cutting or healing of the chain, has stimulated extensive investigations of conductance, resonant transmission, and other impurity-related phenomena in Luttinger liquids \cite{matveev1993tunneling,fendley1995exact,fendley1995exact2,safi1997conductance,ponomarenko1995renormalization}. At the same time, impurity-induced modifications of the single-particle spectrum provide an independent and complementary probe of the underlying many-body correlations. In particular, the local or dynamical density of states exhibits characteristic interaction-dependent power laws and is strongly modified in the vicinity of an impurity \cite{Egger1995friedel1,grishin2004functional,von1998bosonization}.

A major difficulty in treating these problems analytically is the simultaneous presence of arbitrary impurity strength and arbitrary interaction strength. Conventional bosonization is particularly well suited to translationally invariant systems, whereas the introduction of a spatially varying potential generally requires additional approximations or techniques \cite{das2019nonchiral}. Renormalization-group approaches have provided important results in the weak-impurity and weak-interaction regimes \cite{matveev1993tunneling,kane1992transport}, while other approaches have addressed particular impurity configurations or limiting cases. A framework capable of treating strongly inhomogeneous Luttinger liquids while retaining analytical control over the underlying correlation functions is therefore of considerable interest.

Recently, the Unified Field Bosonization Technique (UFBT) was developed to address this class of strongly inhomogeneous Luttinger liquids \cite{das2026unified}. The technique provides a direct Fermi--Bose correspondence adapted to the presence of localized impurity structures and allows the leading-order singular contributions to $N$-point correlation functions to be obtained for arbitrary impurity and interaction strengths. In contrast to approaches that require an expansion in the impurity strength or an RG treatment of the impurity, UFBT starts from the correlation functions of the corresponding inhomogeneous noninteracting problem and incorporates the effect of mutual interactions through the bosonized density correlations. This provides closed analytical expressions for the leading-order singular contributions to $N$-point correlation functions, while retaining the spatial information associated with an arbitrary localized impurity configuration. The technique was subsequently applied to investigate impurity-induced Friedel oscillations and Wigner-crystal-like density correlations in inhomogeneous Luttinger liquids \cite{das2026impurity}.

The availability of these correlation functions makes it possible to investigate several experimentally relevant properties within a common framework. In particular, the two-point fermionic correlation function determines the dynamical density of states, providing access to the energy-resolved single-particle spectrum. Its non-local counterpart describes fermionic propagation between spatially separated points across an inhomogeneous region and provides the basis for analysing the cross-barrier tunneling response. These two observables thus provide complementary probes of the underlying single-particle correlations: the dynamical density of states characterizes the local spectral response, while the cross-barrier tunneling response probes single-particle propagation across the impurity region.

In the present work, these two manifestations of the single-particle correlations are investigated for an inhomogeneous Luttinger liquid containing a localized cluster of impurities using UFBT. The dynamical density of states is obtained from the local fermionic two-point correlation function, and its dependence on interaction strength, position, and energy is examined. The non-local correlation function is then used to investigate the cross-barrier tunneling response, elucidating the influence of impurity scattering and Luttinger-liquid correlations on fermionic propagation across the inhomogeneous region. This unified treatment demonstrates how the analytical correlation functions provided by UFBT connect the many-body description of an inhomogeneous Luttinger liquid with its single-particle spectral and tunneling properties.


\section{System Description}

Consider a one-dimensional interacting fermionic system in the presence of a spatially localized impurity structure. The system is described by the Hamiltonian
\begin{equation}
H=H_0+H_{\mathrm{imp}}+H_{\mathrm{int}}
\label{hamiltonian}
\end{equation}
where $H_0$ describes the free fermions, $H_{\mathrm{imp}}$ describes the localized impurity cluster (localized around an origin $x=0$) and $H_{\mathrm{int}}$ represents the electron-electron interaction given by
\begin{equation}
\begin{aligned}
H_{\mathrm{int}} =\frac{1}{2}\int_{-\infty}^{\infty}dx \int_{-\infty}^{\infty}dx'\,v(x-x')\rho(x)\rho(x'),
\end{aligned}
\label{Hint}
\end{equation}
where $\rho(x)=\psi^\dagger(x)\psi(x)$ is the fermion density. The formalism does not require a particular form of this potential; a single point scatterer, a collection of point impurities, or a finite barrier or well can all be accommodated within the same description. Thus, the spatial inhomogeneity is confined to a finite region, while the system approaches a homogeneous Luttinger liquid away from the impurity structure. The analysis is carried out in the RPA limit, in which the fermionic dispersion is linearized in the vicinity of the Fermi points. More precisely, $k_F$ and $m$ are taken to infinity while their ratio is kept fixed, $\frac{k_F}{m}=v_F$, with $v_F$ the Fermi velocity. The single-particle dispersion near the Fermi surface then takes the form $E\simeq E_F+pv_F$,
rather than the quadratic dispersion of the original microscopic Hamiltonian. Units with $\hbar=1$ are used throughout. Since the impurity region remains physically localized in this limit, its characteristic size $2a$ is scaled such that $2ak_F<\infty$ as $k_F\rightarrow\infty$. The strengths of the barriers or wells are similarly scaled relative to the Fermi energy,
$
E_F=\frac{1}{2}mv_F^2,
$
so that their dimensionless strengths remain finite in the RPA limit.

For a localized impurity potential, the corresponding single-particle scattering problem is characterized by the reflection ($R$) and transmission ($T$) amplitudes at the Fermi energy, satisfying $|R|^2+|T|^2=1$. These amplitudes encode the dependence on the detailed spatial structure of the impurity cluster and provide a convenient means of incorporating its scattering properties into the subsequent many-body analysis. The general formulation therefore does not require a specific impurity geometry, allowing the effect of an arbitrary localized impurity cluster to be retained through its reflection and transmission amplitudes.

The mutual interaction is restricted to short-ranged forward scattering. It is represented in momentum space by
\begin{equation}
v(x-x')=\frac{1}{L}\sum_qv_q\,e^{-iq(x-x')},
\label{interaction}
\end{equation}
where
\(v(x-x')=\frac{1}{L}\sum_q v_q\exp[-iq(x-x')]\) represents the electron-electron interaction, with \(v_q=v_0\) for \(|q|<\Lambda\) and \(v_q=0\) otherwise, where \(\Lambda\ll k_F\) and \(L\) is the system size. Thus, only momentum transfers close to the Fermi points are retained. The temperature is taken to be sufficiently small compared with the interaction bandwidth, $T\ll \Lambda v_F$, so that the low-energy Luttinger-liquid description remains applicable.
\section{Unified Field Bosonization Technique}

The Unified Field Bosonization Technique (UFBT) \cite{das2026unified} provides a framework for describing fermionic excitations in the presence of an impurity region. The essential feature of UFBT is that the fermionic field is constructed from density and current fluctuations on both sides of the impurity region. At low energies, the fermionic field is decomposed into right and left moving components with rapidly oscillating Fermi-momentum factors and slowly varying envelope fields, according to
\begin{equation}
\psi(x)=e^{ik_Fx}\psi_R(x)+e^{-ik_Fx}\psi_L(x).
\label{fermiondecomposition}
\end{equation}
For a branch labelled by $\nu=+1$ for right movers and $\nu=-1$ for left movers, the UFBT prescription takes the form
\begin{equation}
\psi_\nu(x,\sigma,t)\sim
\exp\left[
\frac{i}{\sqrt{2}}
\left(
\theta_\nu(x,\sigma,t)
+\theta_\nu(-x,\sigma,t)
\right)
\right]
\label{ufbt}
\end{equation}
where the phase field is determined by the smooth density fluctuations,
\small
\begin{equation}
\begin{aligned}
\theta_{\nu}(x,\sigma,t) = \pi \int^{x}_{sgn(x)\infty}& dy \bigg( \nu  \mbox{  } \rho_s(y,\sigma,t) \\
&- \int^{y}_{sgn(y)\infty} dy^{'} \mbox{ }\partial_{v_F t }  \mbox{ }\rho_s(y^{'},\sigma,t) \bigg)
\end{aligned}
\end{equation}\normalsize
Here $\rho_s$ denotes the slowly varying component of the density. The appearance of both $x$ and $-x$ in equation (\ref{ufbt}) incorporates the reflection-related structure of the inhomogeneous system directly into the fermionic correlation functions.

Equation (\ref{ufbt}) is to be understood as a Fermi-Bose correspondence for the construction of correlation functions and not as an exact operator identity. Consequently, coefficients multiplying the resulting fermionic correlation functions (prefactors) are fixed independently from the corresponding non-interacting scattering problem. This separation of impurity scattering from interaction effects is useful because the impurity configuration can first be treated at the single-particle level, while the effects of interactions are subsequently introduced through the density correlations.

The calculation therefore proceeds in two stages. First, the density-density correlations of the non-interacting inhomogeneous system are determined using the reflection and transmission amplitudes of the impurity cluster. These correlations retain the spatial information associated with the localized scatterer. Second, the forward-scattering interaction is incorporated into the density correlations and the resulting quantities are inserted into the UFBT prescription in equation (\ref{ufbt}). In this way, the impurity dependence is retained without expanding in the impurity strength.

For the short-ranged interaction specified in equation (\ref{interaction}), the density fluctuations separate into charge and spin sectors. Their characteristic velocities are
\begin{equation}
v_h=\sqrt{v_F^2+\frac{2v_Fv_0}{\pi}},
\qquad
v_n=v_F,
\label{velocities}
\end{equation}
where $v_h$ and $v_n$ denote the holon and spinon velocities, respectively while $v_0$ is the strength of mutual interactions. The corresponding Luttinger liquid interaction parameter can be written as ($c$ for charge and $s$ for spin)
\begin{equation}
K_c=\frac{v_F}{v_h},
\qquad K_s=1.
\label{Kparameters}
\end{equation}
The equality $K_s=1$ follows from the absence of spin-dependent forward-scattering interactions in the present model. The charge and spin sectors consequently propagate with different velocities when the interaction is finite (spin-charge separation). For consistency with the standard notation in the literature, the charge-sector parameter \(K_c\) is denoted by \(g\) in the rest of this work.
Substitution of the interaction-dressed density correlations into equation (\ref{ufbt})  generates the fermionic two-point correlation functions required in the present work. While computing these Green's functions using this prescription, the Gaussian approximation is invoked for the exponential correlation functions. For notational convenience, define
\begin{equation}
\Theta_\nu(x,\sigma,t)= \frac{1}{\sqrt{2}}\left[\theta_\nu(x,\sigma,t)+\theta_\nu(-x,\sigma,t)\right]
\label{Theta}
\end{equation}
The bosonized version of the two-point functions can be written as
\small
\begin{equation}
\begin{aligned}
<\psi_{\nu}(x,\sigma,t)\psi_{\nu'}^{\dagger}(x',&\sigma,t') \mbox{ }\sim \mbox{ }<e^{i\Theta_{\nu}(x,\sigma,t)}e^{-i\Theta_{\nu'}(x',\sigma,t')}>\\
\sim &\mbox{ }e^{\frac{1}{2}<(i\Theta_{\nu}(x,\sigma,t))^2>}e^{\frac{1}{2}<(-i\Theta_{\nu'}(x',\sigma,t'))^2>}\\
&\mbox{ }e^{<(i\Theta_{\nu}(x,\sigma,t))(-i\Theta_{\nu'}(x',\sigma,t'))>}
\label{bch}
\end{aligned}
\end{equation}
\normalsize
Here the factorization in equation (\ref{bch}) follows from the Gaussian approximation, equivalently from retaining the second-order cumulant in the evaluation of the exponential correlation function. The first two factors involve equal-point contractions and, after proper regularization, their time-independent contributions are absorbed into the prefactors. The remaining cross-contraction contains the non-trivial space and time dependent contribution and determines the power-law structure of the Green's functions. Accordingly, the resulting expressions capture the leading singular contribution to the single-particle correlation functions, while higher connected density cumulants are not retained as they contribute to less singular terms beyond this approximation.

The resulting correlation functions contain both the interaction-dependent power-law structure characteristic of a Luttinger liquid and the spatial dependence produced by the impurity scattering. The resulting expressions constitute the basic input for the observables considered in this work. In particular, the local two-point correlation function determines the dynamical density of states, while its nonlocal counterpart connects fermionic propagation between points situated on opposite sides of the impurity region. The latter provides the basis for analysing the tunneling response across the inhomogeneous region.
Within this construction, no expansion in the impurity strength or in the forward-scattering interaction is required. The impurity information enters through the scattering data of the corresponding non-interacting problem, whereas the interaction dependence is encoded in the dressed density correlations and the associated Luttinger-liquid parameters. The resulting UFBT correlation functions represent the leading-order singular contributions relevant to the single-particle observables investigated in this work.

\section{ Results and Discussion}
\subsection{Dynamical Density of States}

The local dynamical density of states (DDOS) provides a direct measure of the low-energy single-particle spectral response at a specified position. In the present work, it is denoted by \(D_x(\omega)\), where \(D_x(\omega)d\omega\) gives the number of single-particle states per unit length within an energy interval \(\hbar d\omega\) around \(\hbar\omega\), measured relative to the Fermi energy. For a one-dimensional non-interacting Fermi gas, the low-energy density of states is independent of energy and is given by
\begin{equation}
\hspace{1.2in}D(\omega)  =  \frac{1}{ \pi v_F }
\label{METH1}
\end{equation}
A similar energy-independent behavior is obtained for a Fermi liquid in the vicinity of the Fermi energy. In contrast, the density of states of a Luttinger liquid is suppressed or enhanced according to a characteristic power law,
\begin{equation}
\hspace{1.2in}D(\omega)\sim \omega^\alpha ,
\end{equation}
where \(\alpha\) is determined by the interaction strength and satisfies \(\alpha=0\) in the non-interacting limit. For an inhomogeneous interacting system, the corresponding spectral response becomes position dependent. It is therefore useful to formulate the DDOS directly in terms of the single-particle correlation function.
At zero temperature, the local DDOS can be expressed as the Fourier transform of the local fermionic anti-commutator,
\small
\begin{equation}
D_{ {\bf{x}} }(\omega)  = \int\limits^{\infty}_{-\infty} \frac{dt}{2\pi} \mbox{   }
     e^{ i t ( \omega  + E_F)  } \mbox{  } \langle \{ \psi({\bf{x}},\sigma,t), \psi^{\dagger}({\bf{x}},\sigma,0) \} \rangle
\label{Local}
\end{equation}
\normalsize
The decomposition of the fermionic field into right ($R$) and left ($L$) moving components separates the local spectral response into non-oscillating and \(2k_F\)-oscillating contributions. Taking $\omega$ relative to the Fermi level, the expression becomes explicitly,
\footnotesize
\begin{equation}
\begin{aligned}
D_{ x }(\omega)
 =& \int^{\infty}_{-\infty} \frac{dt}{2\pi} \mbox{   } e^{ i t \omega  }\mbox{  } \Big( \langle \{ \psi_R(x,\sigma,t), \psi^{\dagger}_R(x,\sigma,0) \} \rangle\\
&\hspace{1cm} +  \langle \{ \psi_L(x,\sigma,t), \psi^{\dagger}_L(x,\sigma,0) \} \rangle \Big)\\
+  &\int^{\infty}_{-\infty} \frac{dt}{2\pi} \mbox{   } e^{ i t  \omega  }\mbox{  } \Big( e^{ 2 i k_F x} \mbox{   }\langle \{ \psi_R(x,\sigma,t), \psi^{\dagger}_L(x,\sigma,0) \} \rangle   \\
&\hspace{1cm}+e^{ -2 i k_F x} \mbox{   } \langle \{ \psi_L(x,\sigma,t), \psi^{\dagger}_R(x,\sigma,0) \} \rangle \Big)
\label{Dx}
\end{aligned}
\end{equation}
\normalsize
The first term in equation (\ref{Dx}) contains the same-chirality (\(RR\) and \(LL\)) two-point functions and gives the non-oscillating or slow contribution to the DDOS. The remaining term contains the mixed-chirality (\(RL\) and \(LR\)) correlators and are accompanied by the factors \(e^{\pm2ik_Fx}\); these constitute the \(2k_F\) Friedel-oscillatory contribution \cite{das2020friedel}. The present analysis is restricted to the non-oscillating part of the local DDOS. For spin-\(\frac12\) fermions, the calculation may be performed for one spin projection, with the final result multiplied by two.
Using the UFBT \cite{das2026unified}, the same-chirality time-ordered two-point function for fermions with spin projection \(\sigma\) can be written as follows, where \(\nu=+1\) and \(\nu=-1\) denote right-moving and left-moving fermions, respectively.
\scriptsize
\begin{equation*}
\begin{aligned}
&\Big\langle T\mbox{ }\psi_{\nu}(x_1,\sigma_1,t_1)\psi_{\nu}^{\dagger}(x_2,\sigma_2,t_2)\Big\rangle \sim
\frac{\delta_{\sigma_1,\sigma_2}}{[(x_1+x_2)^2 -v_h^2 \tau_{12}^2]^{X}}\times\\
& \frac{(4x_1x_2)^{X}}{ (\nu (x_1-x_2) -v_h \tau_{12})^{P} (-\nu (x_1-x_2) -v_h \tau_{12})^{Q} (\nu (x_1-x_2) -v_F \tau_{12})^{\frac{1}{2}}}\
\end{aligned}
\end{equation*}
\normalsize
where \(\tau_{12}=t_1-t_2\) and $\delta_{1,2}$ is the Kronecker delta function. The correlation-function exponents are
\small
\begin{equation}
P=\frac{(v_h+v_F)^2}{8v_hv_F},
\mbox{ }
Q=\frac{(v_h-v_F)^2}{8v_hv_F},
\mbox{ }
X=\frac{v_h^2-v_F^2}{8v_hv_F}.
\label{exponents}
\end{equation}
\normalsize
The position dependence of the DDOS follows from evaluating the above correlation function at the same spatial point, \(x_1=x_2=x\). Apart from phase factors associated with the time-ordering prescription, its scaling form then becomes
\begin{equation}
G_\nu(x,t)\sim\frac{(4x^2)^X}{(4x^2-v_h^2t^2)^X\,t^{P+Q+1/2}} .
\label{DDOSlocalGF}
\end{equation}
Equation (\ref{DDOSlocalGF}) makes the origin of the position-dependent DDOS exponent transparent. The Fourier integral is controlled by times of order \(t\sim\omega^{-1}\). Consequently, the relative importance of the two length scales, \(2|x|\) and \(v_h/\omega\), is characterized by the dimensionless variable
\begin{equation}
\hspace{1in}
\xi=\frac{2\omega |x|}{v_h}.
\label{xi}
\end{equation}
The DDOS can therefore be expressed in the scaling form
$
D_\xi(\omega)\sim\omega^{\alpha(\xi)}.
$
Two limiting regimes follow directly from the correlation function.
For \({\xi\ll1}\), the characteristic Fourier time satisfies \(v_h|t|\gg2|x|\). In this limit, the first factor of the denominator in equation (\ref{DDOSlocalGF}) scales as \small
$
(4x^2-v_h^2t^2)^{X}
\sim
(v_h^2t^2)^{X}. 
$\normalsize 
 \mbox{ }Thus,
\begin{equation}
G_\nu(x,t)\sim
t^{-\left(2X+P+Q+\frac12\right)}.
\end{equation}
Since the Fourier transform of a correlation function behaving as \(t^{-\gamma}\) scales as \(\omega^{\gamma-1}\), the corresponding DDOS exponent is
\begin{equation}
\alpha_{\rm near}=2X+P+Q-\frac12  =
\frac{v_h-v_F}{2v_F}
\end{equation}
In the opposite limit, \({\xi\gg1}\), the characteristic Fourier time obeys \(v_h|t|\ll2|x|\). The spatial factor in equation (\ref{DDOSlocalGF}) then becomes approximately independent of time,
$
(4x^2-v_h^2t^2)^{X}
\sim
(4x^2)^{X},
$
which cancels the corresponding factor \((4x^2)^X\) in the numerator. The remaining time dependence is therefore
\begin{equation}
G_\nu(x,t)\sim
t^{-\left(P+Q+\frac12\right)}.
\end{equation}
The DDOS consequently behaves as
\begin{equation}
D_\xi(\omega)
\sim
\omega^{P+Q-\frac12},
\end{equation}
giving
\begin{equation}
\alpha_{\rm far}
=
P+Q-\frac12
=
\frac{(v_h-v_F)^2}{4v_Fv_h}.
\end{equation}
Thus, the asymptotic DDOS exponent takes the position-dependent limiting form
\begin{equation}
\alpha(\xi)=
\begin{cases}
\dfrac{v_h-v_F}{2v_F},& 0<\xi\ll1,\\[8pt]
\dfrac{(v_h-v_F)^2}{4v_Fv_h},& \xi\gg1.
\end{cases}
\label{alpha}
\end{equation}
These limiting forms describe the crossover between the local spectral response near the impurity cluster and the bulk-like response far from it. Their physical significance and correspondence with established Luttinger-liquid results are discussed separately below. 

At finite temperature, the thermal energy scale \(k_BT\) provides a natural estimate of the energy scale governing the spatial crossover. Here, \(\omega\) denotes the frequency measured relative to the Fermi energy, so that \(\omega=0\) corresponds to the Fermi energy. Replacing the characteristic frequency \(\omega\) in \(\xi=2\omega|x|/v_h\) by \(k_BT/\hbar\) gives the thermal scaling variable
$
\xi_T=2k_BT|x|/(\hbar v_h).
$
Accordingly, the expected low-temperature scaling of the DDOS at the Fermi energy is
\begin{equation}
\hspace{1.0in}D_{\xi_T}(\omega=0,T)\sim T^{\alpha(\xi_T)},
\end{equation}
where \(\alpha(\xi_T)\) approaches \(\alpha_{\rm near}\) for \(\xi_T\ll1\) and \(\alpha_{\rm far}\) for \(\xi_T\gg1\). This relation expresses the expected thermal scaling associated with the two spatial regimes; its quantitative derivation requires the finite-temperature Green's function.

\subsubsection{Far away from the impurity}
It is useful to express the interaction dependence in terms of the Luttinger parameter
$
g=v_F/v_h.
$
In the far field regime, \(\xi\gg1\), equation (\ref{alpha}) gives
\begin{equation}
\alpha_{\rm far}= \frac{(v_h-v_F)^2}{4v_Fv_h}=\frac{1}{4}\left(g+\frac{1}{g}-2\right).
\end{equation}
This is precisely the standard bulk density-of-states exponent for a spinful homogeneous Luttinger liquid; see, for example, Giamarchi \cite{giamarchi2004quantum}, equation (7.27). The exponent is non-negative for both repulsive \((g<1)\) and attractive \((g>1)\) interactions, vanishing only in the non-interacting limit (g=1). Consequently, the low-energy density of states is suppressed relative to that of a uniform non-interacting Fermi gas for either type of interactions. The suppression is therefore not determined simply by whether the interaction is repulsive or attractive, but by the deviation of $g$ from unity.

More explicitly, the exponent is invariant under the transformation \(g\rightarrow 1/g\), so interaction parameters (g) and (1/g) yield the same far-field power-law exponent. For weak interactions, writing \(g=1+\delta\), with \(|\delta|\ll1\), gives \(\alpha_{\rm far}\simeq\delta^2/4\), showing that the leading suppression is quadratic in the deviation from the non-interacting limit. As the interaction strength increases and (g) moves further away from unity, the exponent increases, producing a stronger low-energy suppression.

The recovery of the standard bulk exponent is particularly significant: sufficiently far from the localized impurity cluster, the DDOS exhibits the same interaction-dependent power law as that of a homogeneous Luttinger liquid. The impurity cluster can affect the amplitude and spatial structure of the DDOS, but it does not modify this asymptotic exponent. In the non-interacting limit, \(\alpha_{\rm far}=0\), and the uniform Fermi-gas result in equation (\ref{METH1}) is recovered.

\subsubsection{Near the impurity}

The near-impurity regime, \(0<\xi\ll1\), is characterized by
\begin{equation}
\alpha_{\rm near}
=
 \frac{v_h-v_F}{2v_F}
=
\frac{1}{2}\left(\frac{1}{g}-1\right).
\end{equation}
This result agrees exactly with the established open-boundary exponent for a spinful Luttinger liquid. In particular, Kane and Fisher \cite{kane1992transport} obtained, for spinless fermions, the end density of states \(\rho_{\rm end}(\epsilon)\sim\epsilon^{1/g-1}\). The corresponding exponent for spinful fermions is one half of this value \cite{das2018quantum}, giving
\begin{equation}
\hspace{1in}D_{\rm end}(\omega)\sim\omega^{\frac12(\frac1g-1)},
\end{equation}
which coincides with the present UFBT result.
\begin{figure}[h]
\begin{center}
\includegraphics[scale=0.4]{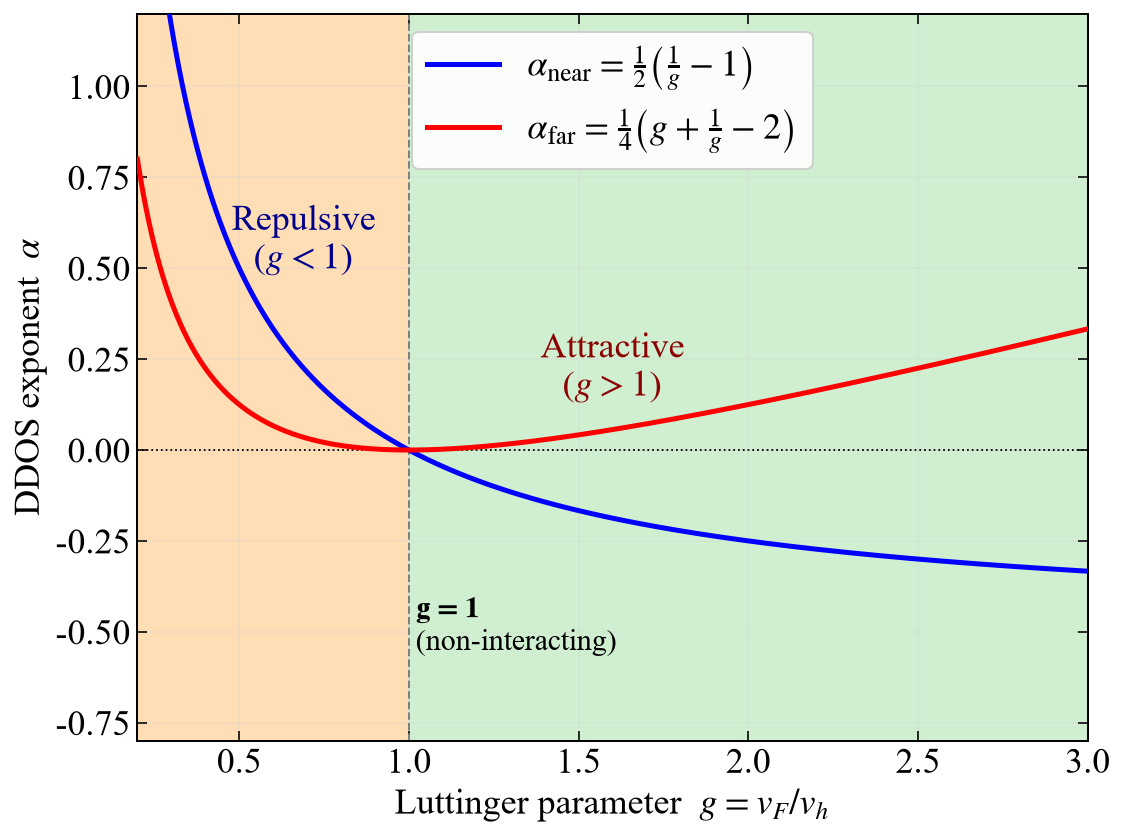}
\end{center}
\caption{Dynamical Density of states exponent as a function of the luttinger liquid parameter $g=v_F/v_h$ both for near and far away from the impurity.}
\label{DDOSexponent}
\end{figure}
Figure (\ref{DDOSexponent}) compares the DDOS exponents near the impurity and sufficiently far from it as functions of the interaction parameter $g$. For repulsive interactions \((g<1)\), both exponents are positive, indicating a power-law suppression of the low-energy DDOS. The near-impurity exponent is larger than the far-impurity exponent throughout this regime, demonstrating that the suppression is stronger near the impurity. This behavior is consistent with the cutting-of-the-chain picture of Kane and Fisher \cite{kane1992transport} in which an impurity effectively separates a one-dimensional interacting system into two segments for repulsive interactions.

At \(g=1\), both exponents vanish, recovering the non-interacting limit. For attractive interactions \((g>1)\), the far-from-impurity exponent remains positive, whereas the near-impurity exponent becomes negative, indicating low-energy enhancement near the impurity. Thus, the impurity modifies the local spectral response differently on the repulsive and attractive sides, while sufficiently far from the impurity the DDOS recovers the standard bulk Luttinger-liquid exponent. Notably, the near-impurity exponent is independent of the reflection amplitude. It follows directly from the UFBT correlation function for the localized impurity structure considered here, without requiring the limiting procedure of setting the reflection coefficient \(|R|^2\rightarrow1\).

\begin{figure}[h]
\begin{center}
\includegraphics[scale=0.5]{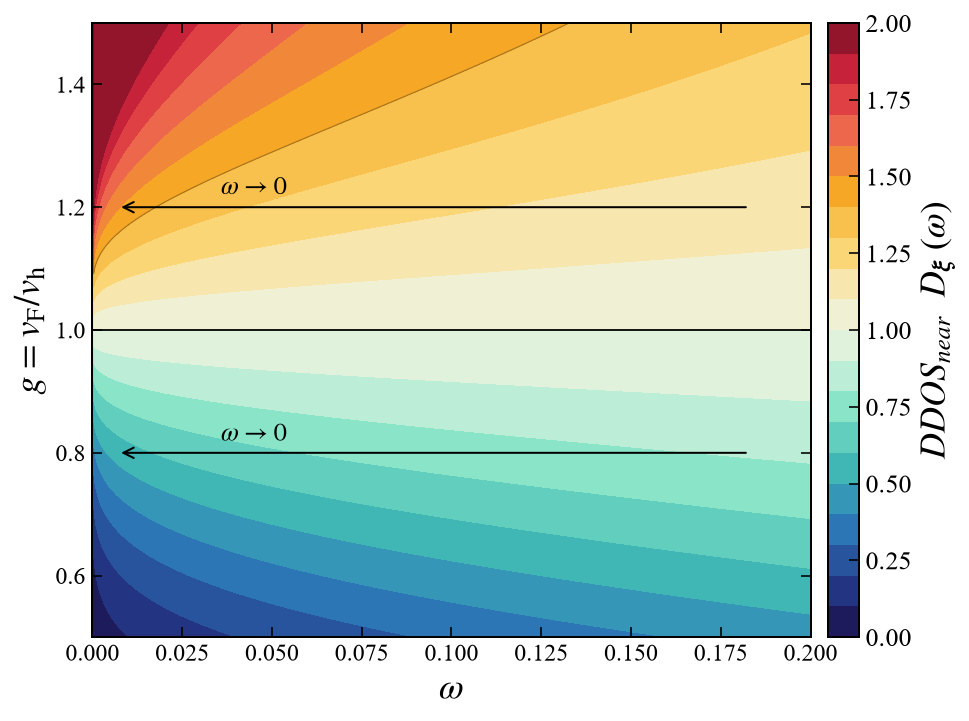}\\
(a)\\
\includegraphics[scale=0.5]{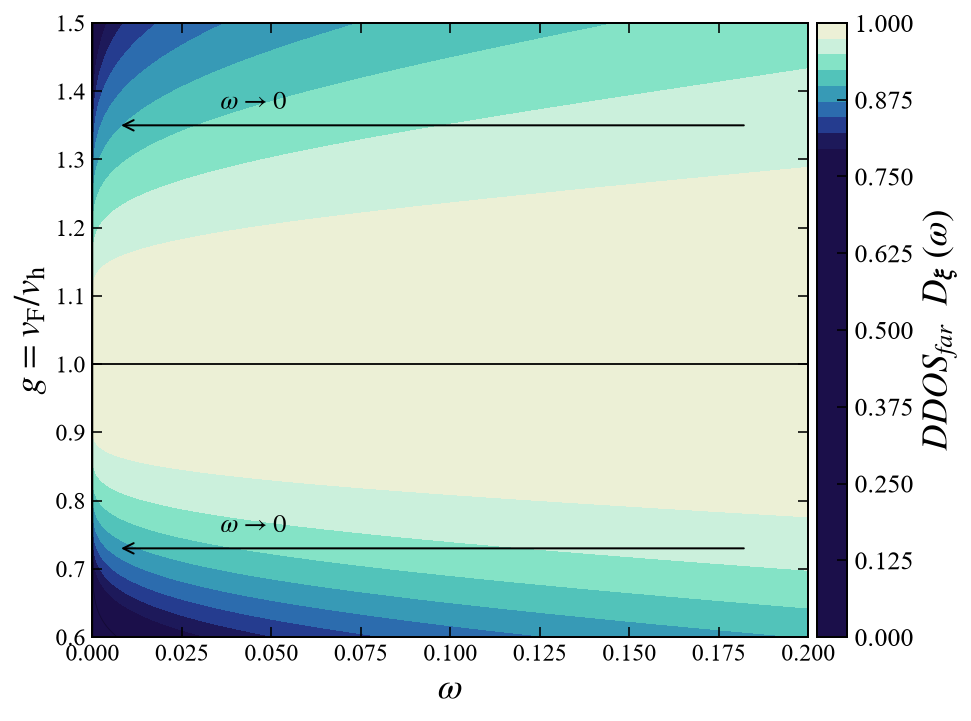}\\
(b)\\
\end{center}
\caption{Dynamical Density of states as a function of energy $\omega$ and the luttinger liquid parameter $g=v_F/v_h$  for (a) near the impurity and (b) far away from the impurity.}
\label{DDOSfarnear}
\end{figure}

The scenarios is further elucidated in Figures (\ref{DDOSfarnear}) where the Density of states $D_\xi(\omega)$ is shown as a function of both energy $\omega$ and Luttinger liquid parameter $g$. It is clearly visible that for $g=1$ the DDOS settled to unity, indicating the exponent becoming zero. That is the DDOS is actually equal to $1/\pi v_F$, here we have chosen a value of $v_F$ such that $\pi v_F =1$. For near the impurity case in Figure (\ref{DDOSfarnear}a), it shows as that as $\omega$ approaches zero (as indicated by the black arrows) the DDOS is suppressed for repulsive interactions and enhanced for attractive interactions. On the other hand, for far away from the impurity in Figure (\ref{DDOSfarnear}b), the maximun value of DDOS is for $g=1$ and for both attractive and repulisve interaction, the DDOS is suppressed as shown by the black arrows. However in this case, the suppresssion is weaker compared to the near the impurity case.

\subsubsection{Comparison with established results}

Further validation is obtained by considering the spinless limit. The exponents in equation (\ref{alpha}) are doubled upon removing the spin degeneracy. Hence,

\begin{equation*}
\begin{aligned}
\alpha_{\rm spinless}(\xi)=
\begin{cases}
\dfrac{1}{g}-1,
&0<\xi\ll1,\\
\dfrac{1}{2}\left(g+\dfrac{1}{g}\right)-1,
&\xi\gg1.
\end{cases}
\end{aligned}
\end{equation*}
The near-impurity exponent agrees with the established boundary exponent for a spinless Luttinger liquid \cite{kane1992transport}.  In particular, for the exactly solvable case \(g=1/2\), the strong-impurity result becomes
\begin{equation}
D(\omega)\sim\omega,
\end{equation}
in agreement with the exact result of von Delft \textit{et al.} \cite{von1998bosonization} and with the results reported by Fabrizio and Gogolin \cite{fabrizio1997comment} and Furusaki \cite{furusaki1997local}.
The far-field spinless exponent,
\begin{equation}
\alpha_{\rm far}
=
\frac12\left(g+\frac1g\right)-1,
\end{equation}
also coincides with the standard bulk result obtained in previous studies, including the functional-bosonization analysis of Grishin \textit{et al.} \cite{grishin2004functional}.

These agreements provide several independent checks on the UFBT formulation. The bulk limit reproduces the conventional Luttinger-liquid exponent, while the near-impurity limit reproduces the established boundary exponent. Moreover, the spinless \(g=1/2\) result agrees with the known exact solution. Importantly, these limits emerge directly from the UFBT correlation function without introducing a renormalization-group flow or taking an additional reflection coefficient \(|R|^2\rightarrow1\) limit. The resulting exponents are independent of the impurity reflection amplitude, while the impurity dependence is retained in the spatial crossover through the scaling variable \(\xi=2\omega|x|/v_h\).

\subsection{Tunneling Conductance}

The conductance of an interacting one-dimensional system can be characterized in more than one way. In the Kubo formulation, the conductance is obtained as a linear response to an externally applied potential and consequently involves a four-point correlation function of fermionic operators. Alternatively, a tunneling experiment probes the propagation of a single fermion between two spatially separated points and is therefore directly related to a two-point Green function \cite{kane1992transport,matveev1993tunneling}. The latter definition is adopted here.

Consider a fermion injected at the left end of the wire of length $L$ and detected at the right end after a time $t$. The corresponding tunneling conductance is proportional to the square of the effective tunneling amplitude,
\begin{equation}
G \sim \frac{ e^2 }{h }  \mbox{   }
| v_F\int^{\infty}_{-\infty}dt\mbox{   }<\{ \psi_{ R } ( -\frac{L}{2},\sigma,0) ,  \psi^{\dagger}_{ R } (\frac{L}{2},\sigma,t)  \}>
 |^2
 \label{TUNNEL1}
\end{equation}\normalsize
This identification of the conductance with an effective tunneling coefficient in the presence of mutual interactions is closely related to the approach of Matveev \textit{et al.} \cite{matveev1993tunneling}. 
The relevant Green function is an unequal-space, unequal-time correlation function. In contrast to the dynamical density of states, which involves an equal-space Green function, the present correlation function describes the propagation of a fermion from one side of the impurity cluster to the other. Since the two spatial points lie on opposite sides of the impurity region, the $RR$ Green function corresponding to opposite-sign spatial coordinates has to be used, with the bare tunneling amplitude entering as a prefactor.
Here, the forward-scattering interaction is assumed to have infinite bandwidth, with a non-zero interaction over the entire momentum range.
For the conductance, the long-time behavior of this Green function is the quantity of interest. In the limit $t\rightarrow\infty$, the terms in the $RR$ Green function that contain the time-dependent factors determine the asymptotic scaling. Upon setting
$x_1=-L/2$ and $x_2=L/2+\epsilon$ and subsequently taking $\epsilon\rightarrow0$, the time-independent factors and the terms that do not contribute to the long-time scaling can be separated from the power-law time dependence. The resulting asymptotic form can be written schematically as
\begin{equation}
\langle
\{
\psi_R(-\frac{L}{2},\sigma,0),
\psi_R^\dagger(\frac{L}{2},\sigma,t)
\}
\rangle
\sim t^{-\gamma},
\qquad t\rightarrow\infty ,
\label{GRRlong}
\end{equation}
where the exponent $\gamma$ is determined by the interaction-dependent exponents appearing in the UFBT Green function. The time integration in equation (\ref{TUNNEL1}) contributes an additional power of $t$, so that the tunneling amplitude scales as $t^{1-\gamma}$. Consequently, upon squaring the amplitude, the conductance acquires the power-law dependence on the characteristic time scale. Using the UFBT Green function derived in Ref.~\cite{das2026unified}, this dependence gives
\begin{equation}
G\sim t^{-(4X+4Q)}\equiv t^{-\eta},
\qquad
\eta=4X+4Q .
\label{Gtime}
\end{equation}
where $X$ and $Q$ are given in equation (\ref{exponents}).
The long-time scale is related to the low-frequency response through
$
t\sim1/\omega.
$
Therefore, the corresponding frequency dependence of the tunneling conductance is
\begin{equation}
G(\omega)\sim \omega^{\eta},
\qquad
\eta=4X+4Q .
\label{Gomega}
\end{equation}
For the d.c. response, the characteristic energy scale is set by temperature through
$
\hbar\omega\sim k_B T .
$
It follows that the low-temperature tunneling conductance scales as
\begin{equation}
G(T)\sim (k_B T)^{\eta},
\qquad
\eta=4X+4Q .
\label{Cond}
\end{equation}
Thus, the temperature dependence of the tunneling conductance is determined directly by the long-time asymptotic behavior of the unequal-space, unequal-time $RR$ Green function. In particular, the impurity cluster enters through the corresponding scattering amplitudes and modifies the spatial structure and amplitude of the Green function, while the power-law exponent is determined by the interaction-dependent exponents.

It is instructive to compare this result with the well-known Kane-Fisher description of a weak link \cite{kane1992transport}. In the weak-link limit, the system approaches two semi-infinite wires connected by a weak tunneling amplitude. For spinless fermions, Kane and Fisher obtain
\begin{equation}
G_{\mathrm{weak-link}}^{\mathrm{spinless}}
\sim (k_B T)^{\frac2g-2},
\end{equation}
where $g$ is the Luttinger parameter ($v_F/v_h$). For spin-$1/2$ fermions, the spin sector remains non-interacting, and the corresponding exponent is reduced by a factor of two,
\begin{equation}
G_{\mathrm{weak-link}}^{\mathrm{spinful}}
\sim (k_B T)^{\frac1g-1}.
\label{KFspinful}
\end{equation}
In the weak-link limit, the exponent of tunneling conductance obtained from the UFBT Green function satisfies
$4X+4Q=1/g-1$.
Consequently, the tunneling conductance obtained from the present approach becomes
\begin{equation}
{G^{UFBT}_{\mathrm{weak-link}}\sim(k_B T)^{4X+4Q}
=(k_B T)^{1/g-1}},
\label{UFBTKF}
\end{equation}
which is precisely the Kane-Fisher result for spin-$1/2$ fermions. This agreement provides a consistency check on the long-time tunneling result in the limiting case where the impurity effectively cuts the wire.
\begin{figure}[h]
\begin{center}
\includegraphics[scale=0.2]{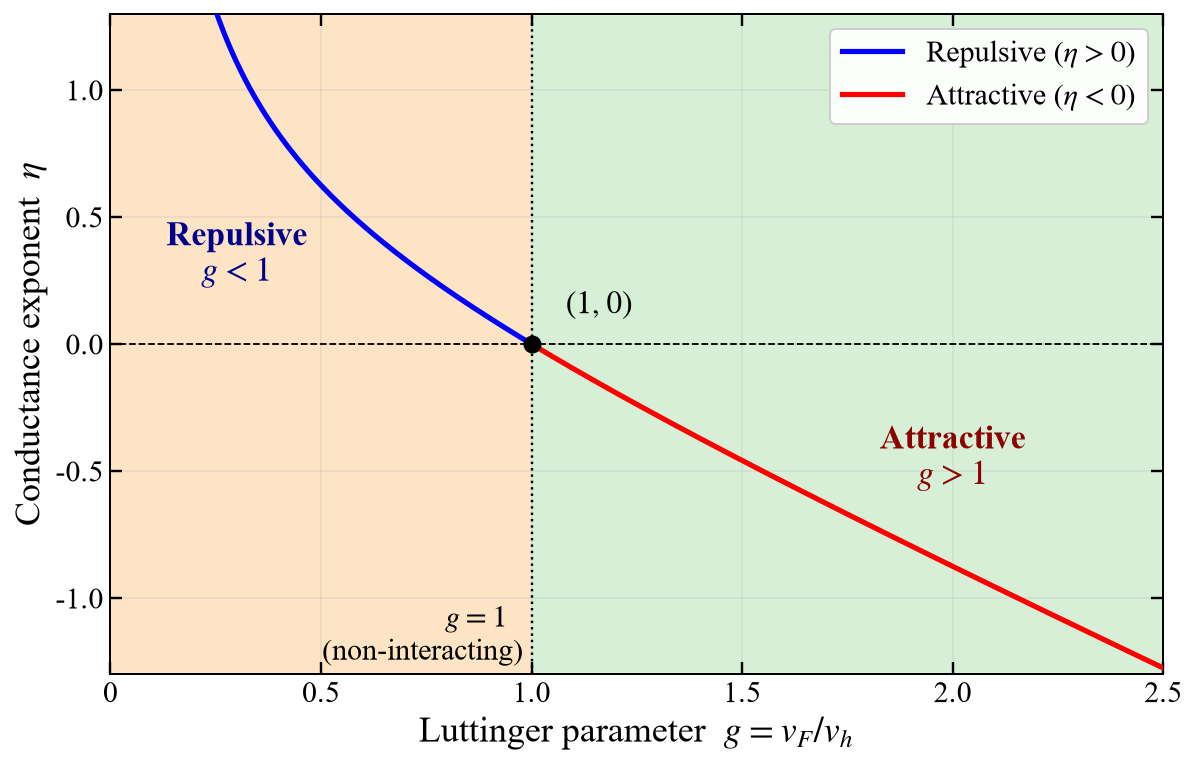}
\end{center}
\caption{Tunneling conductance exponent as a function of the Luttinger liquid parameter $g=v_F/v_h$.}
\label{Gexponent}
\end{figure}

More generally, equation (\ref{Cond}) describes the temperature dependence of the tunneling conductance for a wire without external leads, but in the presence of a localized cluster of barriers or wells and forward-scattering interactions. Figure (\ref{Gexponent}) shows the variation of the tunneling-conductance exponent \(\eta\) as a function of the Luttinger liquid interaction parameter \(g\). The sign of \(\eta\) determines the low-temperature behaviour of the conductance. For repulsive interactions (\(g<1\)), \(\eta>0\), and the conductance decreases as the temperature is lowered, consistent with the tendency towards the cutting of the chain in the weak-link regime. Conversely, for attractive interactions (\(g>1\)), \(\eta<0\), and the conductance increases upon lowering the temperature, indicating a tendency towards the healing of the chain and the recovery of the clean-wire regime. At \(g=1\), the exponent vanishes, yielding a temperature-independent conductance.

\begin{figure}[h]
\begin{center}
\includegraphics[scale=0.5]{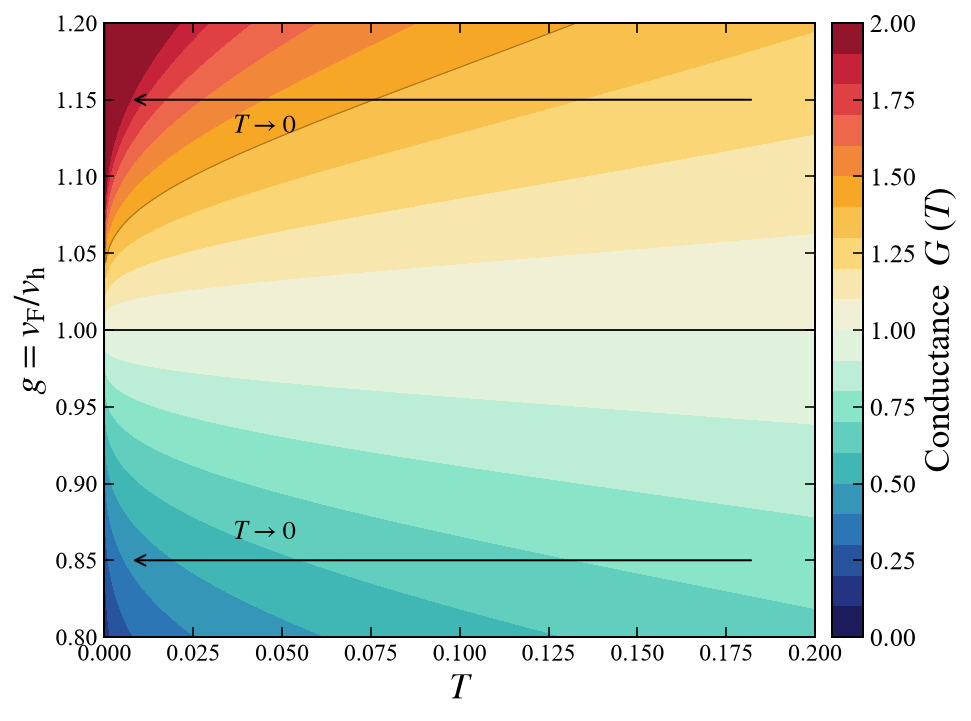}\\
\end{center}
\caption{Tunneling conductance as a function of temperature $T$ and the luttinger liquid parameter $g=v_F/v_h$ }
\label{ConductancePlot}
\end{figure}

These trends are further illustrated in Figure (\ref{ConductancePlot}), which presents the tunneling conductance \(G (T)\) as a function of temperature \(T\) and the interaction parameter \(g\). In the noninteracting limit, \(g=1\), the conductance remains constant at the Landauer value \(\frac{e^2}{h}\mathcal{T}_0\), where \(\mathcal{T}_0\) is the tunneling coefficient of the barrier cluster in the absence of interactions. The conductance has been normalised such that \(\frac{e^2}{h}\mathcal{T}_0=1\) in the plotted results. As indicated by the black arrows, lowering the temperature suppresses the conductance for repulsive interactions and enhances it for attractive interactions. The non-interacting case separates these two contrasting trends, with the conductance remaining unchanged as the temperature varies. Thus, the contour plot illustrates how the interaction strength controls the low-temperature scaling of the tunneling conductance, with repulsive and attractive interactions producing opposite trends relative to the non-interacting limit.
The above result is obtained directly from the asymptotic UFBT Green function and does not require the introduction of a renormalization-group flow of the impurity strength. The crossover between the cutting and healing regimes is therefore encoded in the exponents of the Green function itself.

A useful benchmark for the present result is the renormalization-group analysis of Matveev et al. \cite{matveev1993tunneling}, which treats weak electron-electron interactions. In the present forward-scattering-only treatment, the infinite-bandwidth tunneling conductance is
\begin{equation}
G_{\mathrm{UFBT}}
\sim
\mathcal{T}_0(k_BT)^{4X+4Q},
\end{equation}
where \(\mathcal{T}_0\) is the bare transmission probability which is present in the prefactors of the Green functions (Note that in UFBT the exponents are independent of the transmission parameter). For weak interaction, as considered by Matveev et al., the exponent has the expansion
\begin{equation}
4X+4Q
=
\frac{v_0}{\pi v_F}+O(v_0^2).
\end{equation}
Consequently,
\begin{equation}
G_{\mathrm{UFBT}} = \mathcal{T}\simeq\mathcal{T}_0\left[1+\frac{v_0}{\pi v_F}\ln\left(\frac{k_BT}{E_0}\right)\right],
\end{equation}
to first order in \(v_0\), where \(E_0\) is the reference energy scale.  Define the transmission correction,
\begin{equation}
\Delta\mathcal{T}=\mathcal{T}-\mathcal{T}_0,
\end{equation}
the UFBT result therefore gives
\begin{equation}
\frac{\Delta\mathcal{T}}{\mathcal{T}_0}=\frac{v_0}{\pi v_F}\ln\left(\frac{k_BT}{E_0}\right),
\end{equation}
to first order in \(v_0\). Thus, the UFBT result produces the same logarithmic dependence and the same coefficient \(v_0/(\pi v_F)\) as the weak-interaction correction obtained by Matveev et al. [equation (8) of their manuscript]. The unexpanded UFBT power law, however, retains the full interaction dependence beyond the weak-coupling limit.


\section{Conclusion}
The dynamical density of states (DDOS) and tunneling conductance of an inhomogeneous Luttinger liquid containing a localized cluster of barriers or wells have been investigated within the Unified Field Bosonization Technique (UFBT). For spin-\(\frac{1}{2}\) fermions, the DDOS exhibits distinct near-impurity and far-field power laws, characterized by the exponents \(\alpha_{\mathrm{near}}=(g^{-1}-1)/2\) and \(\alpha_{\mathrm{far}}=(g+g^{-1}-2)/4\), respectively, reproducing the established boundary and bulk Luttinger-liquid results. The low-energy DDOS is suppressed in both regimes for repulsive interactions, more strongly near the impurity cluster, whereas for attractive interactions it is enhanced near the impurity but remains suppressed in the far field. Notably, the near-impurity exponent emerges directly from the UFBT Green function without requiring the strong-impurity limit. The cross-barrier tunneling conductance, derived from the long-time asymptotics of the unequal-space Green function, follows \(G(T)\propto T^{g^{-1}-1}\), reproducing the Kane-Fisher weak-link exponent and predicting a decrease in conductance with decreasing temperature for repulsive interactions and an increase for attractive interactions, consistent with the tendencies towards chain cutting and healing, respectively. In the noninteracting limit, the conductance becomes temperature independent and recovers the Landauer value \((e^2/h)\mathcal{T}_0\), where \(\mathcal{T}_0\) is the bare transmission probability. Furthermore, the weak-interaction expansion reproduces the logarithmic transmission correction obtained by Matveev \textit{et al.} These results establish UFBT as a unified framework in which the asymptotic spectral and transport exponents emerge directly from single-particle correlation functions without an explicit renormalization-group treatment of the impurity strength, while the impurity structure influences the scattering amplitudes and spatial dependence of the responses.\\

\section*{Acknowledgments}

The author gratefully acknowledges Prof. Girish S. Setlur for his guidance and valuable discussions during the development of the ideas underlying this work.


\bibliographystyle{apsrev4-1}
\bibliography{ref}

\begin{thebibliography}{26}%
\makeatletter
\providecommand \@ifxundefined [1]{%
 \@ifx{#1\undefined}
}%
\providecommand \@ifnum [1]{%
 \ifnum #1\expandafter \@firstoftwo
 \else \expandafter \@secondoftwo
 \fi
}%
\providecommand \@ifx [1]{%
 \ifx #1\expandafter \@firstoftwo
 \else \expandafter \@secondoftwo
 \fi
}%
\providecommand \natexlab [1]{#1}%
\providecommand \enquote  [1]{``#1''}%
\providecommand \bibnamefont  [1]{#1}%
\providecommand \bibfnamefont [1]{#1}%
\providecommand \citenamefont [1]{#1}%
\providecommand \href@noop [0]{\@secondoftwo}%
\providecommand \href [0]{\begingroup \@sanitize@url \@href}%
\providecommand \@href[1]{\@@startlink{#1}\@@href}%
\providecommand \@@href[1]{\endgroup#1\@@endlink}%
\providecommand \@sanitize@url [0]{\catcode `\\12\catcode `\$12\catcode
  `\&12\catcode `\#12\catcode `\^12\catcode `\_12\catcode `\%12\relax}%
\providecommand \@@startlink[1]{}%
\providecommand \@@endlink[0]{}%
\providecommand \url  [0]{\begingroup\@sanitize@url \@url }%
\providecommand \@url [1]{\endgroup\@href {#1}{\urlprefix }}%
\providecommand \urlprefix  [0]{URL }%
\providecommand \Eprint [0]{\href }%
\providecommand \doibase [0]{http://dx.doi.org/}%
\providecommand \selectlanguage [0]{\@gobble}%
\providecommand \bibinfo  [0]{\@secondoftwo}%
\providecommand \bibfield  [0]{\@secondoftwo}%
\providecommand \translation [1]{[#1]}%
\providecommand \BibitemOpen [0]{}%
\providecommand \bibitemStop [0]{}%
\providecommand \bibitemNoStop [0]{.\EOS\space}%
\providecommand \EOS [0]{\spacefactor3000\relax}%
\providecommand \BibitemShut  [1]{\csname bibitem#1\endcsname}%
\let\auto@bib@innerbib\@empty
\bibitem [{\citenamefont {Giamarchi}(2004)}]{giamarchi2004quantum}%
  \BibitemOpen
  \bibfield  {author} {\bibinfo {author} {\bibfnamefont {T.}~\bibnamefont
  {Giamarchi}},\ }\href@noop {} {\emph {\bibinfo {title} {Quantum physics in
  one dimension}}}\ (\bibinfo  {publisher} {Clarendon Oxford},\ \bibinfo {year}
  {2004})\BibitemShut {NoStop}%
\bibitem [{\citenamefont {Haldane}(1981)}]{haldane1981luttinger}%
  \BibitemOpen
  \bibfield  {author} {\bibinfo {author} {\bibfnamefont {F.}~\bibnamefont
  {Haldane}},\ }\href@noop {} {\bibfield  {journal} {\bibinfo  {journal}
  {Journal of Physics C: Solid State Physics}\ }\textbf {\bibinfo {volume}
  {14}},\ \bibinfo {pages} {2585} (\bibinfo {year} {1981})}\BibitemShut
  {NoStop}%
\bibitem [{\citenamefont {Bockrath}\ \emph {et~al.}(1999)\citenamefont
  {Bockrath}, \citenamefont {Cobden}, \citenamefont {Lu}, \citenamefont
  {Rinzler}, \citenamefont {Smalley}, \citenamefont {Balents},\ and\
  \citenamefont {McEuen}}]{bockrath1999luttinger}%
  \BibitemOpen
  \bibfield  {author} {\bibinfo {author} {\bibfnamefont {M.}~\bibnamefont
  {Bockrath}}, \bibinfo {author} {\bibfnamefont {D.~H.}\ \bibnamefont
  {Cobden}}, \bibinfo {author} {\bibfnamefont {J.}~\bibnamefont {Lu}}, \bibinfo
  {author} {\bibfnamefont {A.~G.}\ \bibnamefont {Rinzler}}, \bibinfo {author}
  {\bibfnamefont {R.~E.}\ \bibnamefont {Smalley}}, \bibinfo {author}
  {\bibfnamefont {L.}~\bibnamefont {Balents}}, \ and\ \bibinfo {author}
  {\bibfnamefont {P.~L.}\ \bibnamefont {McEuen}},\ }\href@noop {} {\bibfield
  {journal} {\bibinfo  {journal} {Nature}\ }\textbf {\bibinfo {volume} {397}},\
  \bibinfo {pages} {598} (\bibinfo {year} {1999})}\BibitemShut {NoStop}%
\bibitem [{\citenamefont {Bockrath}\ \emph {et~al.}(2001)\citenamefont
  {Bockrath}, \citenamefont {Liang}, \citenamefont {Bozovic}, \citenamefont
  {Hafner}, \citenamefont {Lieber}, \citenamefont {Tinkham},\ and\
  \citenamefont {Park}}]{bockrath2001resonant}%
  \BibitemOpen
  \bibfield  {author} {\bibinfo {author} {\bibfnamefont {M.}~\bibnamefont
  {Bockrath}}, \bibinfo {author} {\bibfnamefont {W.}~\bibnamefont {Liang}},
  \bibinfo {author} {\bibfnamefont {D.}~\bibnamefont {Bozovic}}, \bibinfo
  {author} {\bibfnamefont {J.~H.}\ \bibnamefont {Hafner}}, \bibinfo {author}
  {\bibfnamefont {C.~M.}\ \bibnamefont {Lieber}}, \bibinfo {author}
  {\bibfnamefont {M.}~\bibnamefont {Tinkham}}, \ and\ \bibinfo {author}
  {\bibfnamefont {H.}~\bibnamefont {Park}},\ }\href@noop {} {\bibfield
  {journal} {\bibinfo  {journal} {Science}\ }\textbf {\bibinfo {volume}
  {291}},\ \bibinfo {pages} {283} (\bibinfo {year} {2001})}\BibitemShut
  {NoStop}%
\bibitem [{\citenamefont {Yao}\ \emph {et~al.}(2000)\citenamefont {Yao},
  \citenamefont {Kane},\ and\ \citenamefont {Dekker}}]{yao2000high}%
  \BibitemOpen
  \bibfield  {author} {\bibinfo {author} {\bibfnamefont {Z.}~\bibnamefont
  {Yao}}, \bibinfo {author} {\bibfnamefont {C.~L.}\ \bibnamefont {Kane}}, \
  and\ \bibinfo {author} {\bibfnamefont {C.}~\bibnamefont {Dekker}},\
  }\href@noop {} {\bibfield  {journal} {\bibinfo  {journal} {Physical Review
  Letters}\ }\textbf {\bibinfo {volume} {84}},\ \bibinfo {pages} {2941}
  (\bibinfo {year} {2000})}\BibitemShut {NoStop}%
\bibitem [{\citenamefont {Auslaender}\ \emph {et~al.}(2000)\citenamefont
  {Auslaender}, \citenamefont {Yacoby}, \citenamefont {De~Picciotto},
  \citenamefont {Baldwin}, \citenamefont {Pfeiffer},\ and\ \citenamefont
  {West}}]{auslaender2000experimental}%
  \BibitemOpen
  \bibfield  {author} {\bibinfo {author} {\bibfnamefont {O.}~\bibnamefont
  {Auslaender}}, \bibinfo {author} {\bibfnamefont {A.}~\bibnamefont {Yacoby}},
  \bibinfo {author} {\bibfnamefont {R.}~\bibnamefont {De~Picciotto}}, \bibinfo
  {author} {\bibfnamefont {K.}~\bibnamefont {Baldwin}}, \bibinfo {author}
  {\bibfnamefont {L.}~\bibnamefont {Pfeiffer}}, \ and\ \bibinfo {author}
  {\bibfnamefont {K.}~\bibnamefont {West}},\ }\href@noop {} {\bibfield
  {journal} {\bibinfo  {journal} {Physical Review Letters}\ }\textbf {\bibinfo
  {volume} {84}},\ \bibinfo {pages} {1764} (\bibinfo {year}
  {2000})}\BibitemShut {NoStop}%
\bibitem [{\citenamefont {Yacoby}\ \emph {et~al.}(1997)\citenamefont {Yacoby},
  \citenamefont {Stormer}, \citenamefont {Baldwin}, \citenamefont {Pfeiffer},\
  and\ \citenamefont {West}}]{yacoby1997magneto}%
  \BibitemOpen
  \bibfield  {author} {\bibinfo {author} {\bibfnamefont {A.}~\bibnamefont
  {Yacoby}}, \bibinfo {author} {\bibfnamefont {H.}~\bibnamefont {Stormer}},
  \bibinfo {author} {\bibfnamefont {K.}~\bibnamefont {Baldwin}}, \bibinfo
  {author} {\bibfnamefont {L.}~\bibnamefont {Pfeiffer}}, \ and\ \bibinfo
  {author} {\bibfnamefont {K.}~\bibnamefont {West}},\ }\href@noop {} {\bibfield
   {journal} {\bibinfo  {journal} {Solid State Communications}\ }\textbf
  {\bibinfo {volume} {101}},\ \bibinfo {pages} {77} (\bibinfo {year}
  {1997})}\BibitemShut {NoStop}%
\bibitem [{\citenamefont {Ge}\ \emph {et~al.}(2024)\citenamefont {Ge},
  \citenamefont {Halbinger}, \citenamefont {Lee}, \citenamefont {von Delft},\
  and\ \citenamefont {Kugler}}]{ge2024analytic}%
  \BibitemOpen
  \bibfield  {author} {\bibinfo {author} {\bibfnamefont {A.}~\bibnamefont
  {Ge}}, \bibinfo {author} {\bibfnamefont {J.}~\bibnamefont {Halbinger}},
  \bibinfo {author} {\bibfnamefont {S.-S.~B.}\ \bibnamefont {Lee}}, \bibinfo
  {author} {\bibfnamefont {J.}~\bibnamefont {von Delft}}, \ and\ \bibinfo
  {author} {\bibfnamefont {F.~B.}\ \bibnamefont {Kugler}},\ }\href@noop {}
  {\bibfield  {journal} {\bibinfo  {journal} {Annalen der Physik}\ }\textbf
  {\bibinfo {volume} {536}},\ \bibinfo {pages} {2300504} (\bibinfo {year}
  {2024})}\BibitemShut {NoStop}%
\bibitem [{\citenamefont {Von~Delft}\ and\ \citenamefont
  {Schoeller}(1998)}]{von1998bosonization}%
  \BibitemOpen
  \bibfield  {author} {\bibinfo {author} {\bibfnamefont {J.}~\bibnamefont
  {Von~Delft}}\ and\ \bibinfo {author} {\bibfnamefont {H.}~\bibnamefont
  {Schoeller}},\ }\href@noop {} {\bibfield  {journal} {\bibinfo  {journal}
  {Annalen der Physik}\ }\textbf {\bibinfo {volume} {7}},\ \bibinfo {pages}
  {225} (\bibinfo {year} {1998})}\BibitemShut {NoStop}%
\bibitem [{\citenamefont {Das}\ and\ \citenamefont
  {Setlur}(2019)}]{das2019transport}%
  \BibitemOpen
  \bibfield  {author} {\bibinfo {author} {\bibfnamefont {J.~P.}\ \bibnamefont
  {Das}}\ and\ \bibinfo {author} {\bibfnamefont {G.~S.}\ \bibnamefont
  {Setlur}},\ }\href@noop {} {\bibfield  {journal} {\bibinfo  {journal}
  {Physica E: Low-dimensional Systems and Nanostructures}\ }\textbf {\bibinfo
  {volume} {110}},\ \bibinfo {pages} {39} (\bibinfo {year} {2019})}\BibitemShut
  {NoStop}%
\bibitem [{\citenamefont {Das}\ \emph {et~al.}(2020)\citenamefont {Das},
  \citenamefont {Chowdhury},\ and\ \citenamefont {Setlur}}]{das2020friedel}%
  \BibitemOpen
  \bibfield  {author} {\bibinfo {author} {\bibfnamefont {J.~P.}\ \bibnamefont
  {Das}}, \bibinfo {author} {\bibfnamefont {C.}~\bibnamefont {Chowdhury}}, \
  and\ \bibinfo {author} {\bibfnamefont {G.~S.}\ \bibnamefont {Setlur}},\
  }\href@noop {} {\bibfield  {journal} {\bibinfo  {journal} {Physica Scripta}\
  }\textbf {\bibinfo {volume} {95}},\ \bibinfo {pages} {075710} (\bibinfo
  {year} {2020})}\BibitemShut {NoStop}%
\bibitem [{\citenamefont {Kane}\ and\ \citenamefont
  {Fisher}(1992{\natexlab{a}})}]{kane1992transport}%
  \BibitemOpen
  \bibfield  {author} {\bibinfo {author} {\bibfnamefont {C.}~\bibnamefont
  {Kane}}\ and\ \bibinfo {author} {\bibfnamefont {M.~P.}\ \bibnamefont
  {Fisher}},\ }\href@noop {} {\bibfield  {journal} {\bibinfo  {journal}
  {Physical Review Letters}\ }\textbf {\bibinfo {volume} {68}},\ \bibinfo
  {pages} {1220} (\bibinfo {year} {1992}{\natexlab{a}})}\BibitemShut {NoStop}%
\bibitem [{\citenamefont {Kane}\ and\ \citenamefont
  {Fisher}(1992{\natexlab{b}})}]{kane1992resonant}%
  \BibitemOpen
  \bibfield  {author} {\bibinfo {author} {\bibfnamefont {C.}~\bibnamefont
  {Kane}}\ and\ \bibinfo {author} {\bibfnamefont {M.~P.}\ \bibnamefont
  {Fisher}},\ }\href@noop {} {\bibfield  {journal} {\bibinfo  {journal}
  {Physical Review B}\ }\textbf {\bibinfo {volume} {46}},\ \bibinfo {pages}
  {7268} (\bibinfo {year} {1992}{\natexlab{b}})}\BibitemShut {NoStop}%
\bibitem [{\citenamefont {Matveev}\ \emph {et~al.}(1993)\citenamefont
  {Matveev}, \citenamefont {Yue},\ and\ \citenamefont
  {Glazman}}]{matveev1993tunneling}%
  \BibitemOpen
  \bibfield  {author} {\bibinfo {author} {\bibfnamefont {K.}~\bibnamefont
  {Matveev}}, \bibinfo {author} {\bibfnamefont {D.}~\bibnamefont {Yue}}, \ and\
  \bibinfo {author} {\bibfnamefont {L.}~\bibnamefont {Glazman}},\ }\href@noop
  {} {\bibfield  {journal} {\bibinfo  {journal} {Physical Review Letters}\
  }\textbf {\bibinfo {volume} {71}},\ \bibinfo {pages} {3351} (\bibinfo {year}
  {1993})}\BibitemShut {NoStop}%
\bibitem [{\citenamefont {Fendley}\ \emph
  {et~al.}(1995{\natexlab{a}})\citenamefont {Fendley}, \citenamefont {Ludwig},\
  and\ \citenamefont {Saleur}}]{fendley1995exact}%
  \BibitemOpen
  \bibfield  {author} {\bibinfo {author} {\bibfnamefont {P.}~\bibnamefont
  {Fendley}}, \bibinfo {author} {\bibfnamefont {A.}~\bibnamefont {Ludwig}}, \
  and\ \bibinfo {author} {\bibfnamefont {H.}~\bibnamefont {Saleur}},\
  }\href@noop {} {\bibfield  {journal} {\bibinfo  {journal} {Physical Review
  Letters}\ }\textbf {\bibinfo {volume} {74}},\ \bibinfo {pages} {3005}
  (\bibinfo {year} {1995}{\natexlab{a}})}\BibitemShut {NoStop}%
\bibitem [{\citenamefont {Fendley}\ \emph
  {et~al.}(1995{\natexlab{b}})\citenamefont {Fendley}, \citenamefont {Ludwig},\
  and\ \citenamefont {Saleur}}]{fendley1995exact2}%
  \BibitemOpen
  \bibfield  {author} {\bibinfo {author} {\bibfnamefont {P.}~\bibnamefont
  {Fendley}}, \bibinfo {author} {\bibfnamefont {A.}~\bibnamefont {Ludwig}}, \
  and\ \bibinfo {author} {\bibfnamefont {H.}~\bibnamefont {Saleur}},\
  }\href@noop {} {\bibfield  {journal} {\bibinfo  {journal} {Physical Review
  B}\ }\textbf {\bibinfo {volume} {52}},\ \bibinfo {pages} {8934} (\bibinfo
  {year} {1995}{\natexlab{b}})}\BibitemShut {NoStop}%
\bibitem [{\citenamefont {Safi}(1997)}]{safi1997conductance}%
  \BibitemOpen
  \bibfield  {author} {\bibinfo {author} {\bibfnamefont {I.}~\bibnamefont
  {Safi}},\ }\href@noop {} {\bibfield  {journal} {\bibinfo  {journal} {Physical
  Review-Section B-Condensed Matter}\ }\textbf {\bibinfo {volume} {55}},\
  \bibinfo {pages} {7331} (\bibinfo {year} {1997})}\BibitemShut {NoStop}%
\bibitem [{\citenamefont {Ponomarenko}(1995)}]{ponomarenko1995renormalization}%
  \BibitemOpen
  \bibfield  {author} {\bibinfo {author} {\bibfnamefont {V.}~\bibnamefont
  {Ponomarenko}},\ }\href@noop {} {\bibfield  {journal} {\bibinfo  {journal}
  {Physical Review B}\ }\textbf {\bibinfo {volume} {52}},\ \bibinfo {pages}
  {R8666} (\bibinfo {year} {1995})}\BibitemShut {NoStop}%
\bibitem [{\citenamefont {Egger}\ and\ \citenamefont
  {Grabert}(1995)}]{Egger1995friedel1}%
  \BibitemOpen
  \bibfield  {author} {\bibinfo {author} {\bibfnamefont {R.}~\bibnamefont
  {Egger}}\ and\ \bibinfo {author} {\bibfnamefont {H.}~\bibnamefont
  {Grabert}},\ }\href@noop {} {\bibfield  {journal} {\bibinfo  {journal}
  {Physical Review Letters}\ }\textbf {\bibinfo {volume} {75}},\ \bibinfo
  {pages} {3505} (\bibinfo {year} {1995})}\BibitemShut {NoStop}%
\bibitem [{\citenamefont {Grishin}\ \emph {et~al.}(2004)\citenamefont
  {Grishin}, \citenamefont {Yurkevich},\ and\ \citenamefont
  {Lerner}}]{grishin2004functional}%
  \BibitemOpen
  \bibfield  {author} {\bibinfo {author} {\bibfnamefont {A.}~\bibnamefont
  {Grishin}}, \bibinfo {author} {\bibfnamefont {I.~V.}\ \bibnamefont
  {Yurkevich}}, \ and\ \bibinfo {author} {\bibfnamefont {I.~V.}\ \bibnamefont
  {Lerner}},\ }\href@noop {} {\bibfield  {journal} {\bibinfo  {journal}
  {Physical Review B}\ }\textbf {\bibinfo {volume} {69}},\ \bibinfo {pages}
  {165108} (\bibinfo {year} {2004})}\BibitemShut {NoStop}%
\bibitem [{\citenamefont {Das}\ \emph {et~al.}(2019)\citenamefont {Das},
  \citenamefont {Chowdhury},\ and\ \citenamefont {Setlur}}]{das2019nonchiral}%
  \BibitemOpen
  \bibfield  {author} {\bibinfo {author} {\bibfnamefont {J.~P.}\ \bibnamefont
  {Das}}, \bibinfo {author} {\bibfnamefont {C.}~\bibnamefont {Chowdhury}}, \
  and\ \bibinfo {author} {\bibfnamefont {G.~S.}\ \bibnamefont {Setlur}},\
  }\href@noop {} {\bibfield  {journal} {\bibinfo  {journal} {Theoretical and
  Mathematical Physics}\ }\textbf {\bibinfo {volume} {199}},\ \bibinfo {pages}
  {736} (\bibinfo {year} {2019})}\BibitemShut {NoStop}%
\bibitem [{\citenamefont {P}\ \emph {et~al.}(2026{\natexlab{a}})\citenamefont
  {P}, \citenamefont {M},\ and\ \citenamefont {Das}}]{das2026unified}%
  \BibitemOpen
  \bibfield  {author} {\bibinfo {author} {\bibfnamefont {S.}~\bibnamefont {P}},
  \bibinfo {author} {\bibfnamefont {V.~S.}\ \bibnamefont {M}}, \ and\ \bibinfo
  {author} {\bibfnamefont {J.~P.}\ \bibnamefont {Das}},\ }\href@noop {}
  {\bibfield  {journal} {\bibinfo  {journal} {arXiv preprint arXiv:2609.21973}\
  } (\bibinfo {year} {2026}{\natexlab{a}})}\BibitemShut {NoStop}%
\bibitem [{\citenamefont {P}\ \emph {et~al.}(2026{\natexlab{b}})\citenamefont
  {P}, \citenamefont {M},\ and\ \citenamefont {Das}}]{das2026impurity}%
  \BibitemOpen
  \bibfield  {author} {\bibinfo {author} {\bibfnamefont {S.}~\bibnamefont {P}},
  \bibinfo {author} {\bibfnamefont {V.~S.}\ \bibnamefont {M}}, \ and\ \bibinfo
  {author} {\bibfnamefont {J.~P.}\ \bibnamefont {Das}},\ }\href@noop {}
  {\bibfield  {journal} {\bibinfo  {journal} {arXiv preprint arXiv:2609.40053}\
  } (\bibinfo {year} {2026}{\natexlab{b}})}\BibitemShut {NoStop}%
\bibitem [{\citenamefont {Das}\ and\ \citenamefont
  {Setlur}(2018)}]{das2018quantum}%
  \BibitemOpen
  \bibfield  {author} {\bibinfo {author} {\bibfnamefont {J.~P.}\ \bibnamefont
  {Das}}\ and\ \bibinfo {author} {\bibfnamefont {G.~S.}\ \bibnamefont
  {Setlur}},\ }\href@noop {} {\bibfield  {journal} {\bibinfo  {journal}
  {International Journal of Modern Physics A}\ ,\ \bibinfo {pages} {1850174}}
  (\bibinfo {year} {2018})}\BibitemShut {NoStop}%
\bibitem [{\citenamefont {Fabrizio}\ and\ \citenamefont
  {Gogolin}(1997)}]{fabrizio1997comment}%
  \BibitemOpen
  \bibfield  {author} {\bibinfo {author} {\bibfnamefont {M.}~\bibnamefont
  {Fabrizio}}\ and\ \bibinfo {author} {\bibfnamefont {A.~O.}\ \bibnamefont
  {Gogolin}},\ }\href@noop {} {\bibfield  {journal} {\bibinfo  {journal}
  {Physical review letters}\ }\textbf {\bibinfo {volume} {78}},\ \bibinfo
  {pages} {4527} (\bibinfo {year} {1997})}\BibitemShut {NoStop}%
\bibitem [{\citenamefont {Furusaki}(1997)}]{furusaki1997local}%
  \BibitemOpen
  \bibfield  {author} {\bibinfo {author} {\bibfnamefont {A.}~\bibnamefont
  {Furusaki}},\ }\href@noop {} {\bibfield  {journal} {\bibinfo  {journal}
  {Physical Review B}\ }\textbf {\bibinfo {volume} {56}},\ \bibinfo {pages}
  {9352} (\bibinfo {year} {1997})}\BibitemShut {NoStop}%
\end{thebibliography}%
\normalsize

\end{document}